\documentclass[pre,showpacs,floats,twocolumn,floatfix,superscriptaddress]{revtex4-1}

\usepackage{graphics,graphicx,dcolumn,bm,fleqn,epic,eepic,float}
\usepackage{amssymb,amsmath,multirow,rotate,rotating,color,float}
\usepackage[utf8]{inputenc}

\begin{document}

\title{From Bijels to Pickering emulsions: a lattice Boltzmann study}

\author{Fabian Jansen}
\affiliation{Institute for Computational Physics, University of Stuttgart, Pfaffenwaldring 27, D-70569 Stuttgart, Germany} 

\author{Jens Harting} 
\affiliation{Department of Applied Physics, Eindhoven University of
  Technology, P.\,O. Box 513, 5600\,MB Eindhoven, The Netherlands}
\affiliation{Institute for Computational Physics, University of Stuttgart, Pfaffenwaldring 27, D-70569 Stuttgart, Germany} 

\date{\today}

\begin{abstract}
Particle stabilized emulsions are ubiquitous in the food and cosmetics
industry, but our understanding of the influence of microscopic fluid-particle
and particle-particle interactions on the macroscopic rheology is still
limited. In this paper we present a simulation algorithm based on a
multicomponent lattice Boltzmann model to describe the solvents combined with a
molecular dynamics solver for the description of the solved particles. It is
shown that the model allows a wide variation of fluid properties and arbitrary
contact angles on the particle surfaces. We demonstrate its applicability by
studying the transition from a ``bicontinuous interfacially jammed emulsion
gel'' (bijel) to a ``Pickering emulsion'' in dependence on the contact angle,
the particle concentration, and the ratio of the solvents.
\end{abstract}
 
\maketitle

\section{Introduction}
\label{sec_introduction}
Using particles in a manner similar to surfactants in order to stabilize
emulsions is very attractive in particular for the food-, cosmetics-, and
medical industry to stabilize, e.g. barbecue sauces and sun cremes or in order
to produce sophisticated ways to deliver drugs at the right position in the
human body. 
The microscopic processes leading to the commercial interest can be understood
by assuming an oil-water mixture.  Without any additives, phase separation
would take place and the oil would float on top of the water. Adding small
particles, however, causes these particles to diffuse to the interface which is
being stabilized due to a reduced surface energy. If for example individual droplets of one phase are
covered by particles, such systems are also referred to as
``Pickering emulsions'' and are known since the beginning of the last
century~\cite{bib:ramsden-1903,bib:pickering-1907}. Particularly interesting
properties of such emulsions are the blocking of Ostwald ripening and the
rheological properties due to irreversible particle adsorption at interfaces as
well as interface bridging due to particle monolayers~\cite{bib:arditty-whitby-binks-schmitt-leal-calderon,bib:binks-clint-whitby-2005,bib:arditty-schmitt-giermanska-Kahn-leal-calderon,bib:binks-horozov}. 

Recently, there has been a growing interest in particles suspended in
multiphase or multicomponent
flows~\cite{bib:binks-horozov,bib:binks-2002,bib:herzig-robert-zand-cipelletti-pusey-clegg,bib:arditty-whitby-binks-schmitt-leal-calderon},
which led to the discovery of a new material type, the ``bicontinous
interfacially jammed emulsion gel'' (bijel)~\cite{bib:cates-clegg-2008}.  The
existence of the bijel was predicted in 2005 by Stratford et
al.~\cite{bib:stratford-adhikari-pagonabarraga-desplat-cates-2005,bib:kim-stratford-adhikari-cates} and
experimentally confirmed by Herzig et al. in
2007~\cite{bib:herzig-white-schofield-poon-clegg}. In contrast to Pickering
emulsions which consist of unconnected particle stabilised droplets distributed
in a second continuous fluid phase, the bijel shows an 
interface between two continuous fluid phases which is covered by particles.

Since the particles used for stabilization have a larger size than surfactant
molecules and do not present any amphiphilic properties, concepts developed for
the description of surfactant stabilized systems are often not applicable.
Instead, theoretical models have to be developed and experiments have to be
performed which consider the specific properties of particle-stabilized
systems. These include the particle's contact angle, the strong interparticle
capillary forces, or the pH value and electrolyte concentration of the
solvents~\cite{bib:binks-horozov,bib:binks-2002,bib:tcholakova-denkov-lips-2008,bib:aveyard-binks-clint,bib:arditty-schmitt-giermanska-Kahn-leal-calderon}.
However, even today our quantitative understanding of solid stabilized
emulsions is still far from satisfactory.

Computer simulations are promising to understand the dynamic properties of
particle stabilized multiphase flows. However, the shortcomings of traditional
simulation methods quickly become obvious: a suitable simulation algorithm is
not only required to deal with simple fluid dynamics but has to be able to
simulate several fluid species while also considering the motion of the
particles and the fluid-particle interactions. Some recent approaches trying to
solve these problems utilize the lattice Boltzmann method for the description of
the solvents~\cite{bib:succi-01}.
The lattice Boltzmann method can be seen as an alternative to conventional
Navier Stokes solvers and is well established in the literature. It is
attractive for the current application since a number of multiphase and
multicomponent models exist which are comparably straightforward to implement~\cite{bib:shan-chen-93,bib:shan-doolen,bib:shan-chen-liq-gas,bib:swift-osborn-yeomans,bib:swift-orlandini-osborn-yeomans,bib:gunstensen-rothman-zaleski-zanetti,bib:lishchuk-care-halliday-2003}.
In addition, the method has been combined with a molecular dynamics algorithm
to simulate arbitrarily shaped particles in flow and is commonly used to study
the behavior of particle-laden single phase flows~\cite{Ladd94a,Ladd94b,Ladd01,bib:jens-herrmann-bennaim:2008}.

A few groups combined multiphase lattice Boltzmann solvers with the known
algorithms for suspended
particles~\cite{bib:stratford-adhikari-pagonabarraga-desplat-cates-2005,bib:joshi-sun}.
In this paper we follow an alternative approach: we present a method based on
the multicomponent lattice Boltzmann model of Shan and
Chen~\cite{bib:shan-chen-93} which allows the simulation of multiple fluid
components with surface tension. Our model generally allows arbitrary movements
and rotations of arbitrarily shaped hard shell particles. It does not require
fluid-filled particles and thus does not suffer from unphysical behavior caused
by oscillations of the inner fluid~\cite{bib:onishi-kawasaki-chen-ohashi}.
Further, it allows an arbitrary choice of the particle wettability -- one of
the most important parameters for the dynamics of multiphase
suspensions~\cite{bib:binks-horozov,bib:binks-2002}.

The remainder of the paper is organised as follows: after a description of the
Shan-Chen approach for multicomponent lattice Boltzmann simulations and an
extension of the lattice Boltzmann method to simulate suspensions, a way to
combine the two methods is proposed.
The influence of the parameters of the model on the contact angle as a measure
of wettability is studied in the following section. Then the suitability of the
new method is tested by performing a detailed study of the formation of bijels
and Pickering emulsions. 

\section{The multicomponent lattice Boltzmann model}
\label{sec_multicomponent}
The dynamics of the fluid solvents is simulated by a multicomponent lattice
Boltzmann model following the approach of Shan and
Chen~\cite{bib:shan-chen-93}. Here, each component follows a lattice Boltzmann
equation
\begin{equation}
\label{eq:boltz}
f_i^{c} \left( \mathbf{x} + \mathbf{c}_i, t + 1 \right) = f_i^{c} \left( \mathbf{x}, t \right) + \Omega_i^{c} (\mathbf{x},t),
\end{equation}
where $f_i^{c}\left( \mathbf{x}, t \right)$ is the single particle distribution
function for component $c$ in the direction $\mathbf{c}_i$ ($i=1, \ldots, N$)
at a discrete lattice position $\mathbf{x}$ and at timestep $t$. In this work
we use exclusively the so-called D3Q19 implementation, where
$N=19$ velocities are used on a three dimensional lattice. For simplicity,
the length of a timestep and the lattice constant are set to 1, i.e. all
units are given in lattice units if not stated otherwise.
$\Omega_i^{c}$ is the Bhatnagar-Gross-Krook (BGK) collision operator~\cite{bib:bgk},
  \begin{equation}
   \Omega_i^{c}(\mathbf{x},t) = - \frac{f_i^{c}(\mathbf{x},t)- f_i^{\textnormal{eq}^c}\left(\rho^{c}(\mathbf{x},t),{\bf u}^c({\bf x},t)\right)}{\tau^c},
  \end{equation}
which is a relaxation towards the local equilibrium distribution function
\begin{equation}
\label{Equil}
\!f_i^{\textnormal{eq}^{c}}\!\!\! =\!\!
 \zeta_i\rho^{c}\!\left[\!1\!+\!
\frac{{\bf c}_i {\bf u}}{c_s^2}\! +\! \frac{({\bf c}_i {\bf u})^2}{2c_s^4}
\!-\!\frac{u^2}{2c_s^2}\!+\!\frac{({\bf c}_i {\bf u})^3}{6c_s^6}
\!-\!\frac{u^2({\bf c}_i {\bf u})}{2c_s^4}\right]\!
\end{equation}
on a time scale given by the relaxation time $\tau^c$~\cite{bib:chen-chen-martinez-matthaeus}. 
Here,
$\rho^{c}\left(\mathbf{x},t\right)=\rho_0\sum_i f^{c}_{i}\left(\mathbf{x},t\right)$
is the fluid density with reference density $\rho_0$ and ${\bf u} = {\bf
u}^c({\bf x},t)$ is the macroscopic bulk velocity of the fluid, given by
$\rho^c({\bf x},t){\bf u}^{c}({\bf x},t) \equiv \sum_i f_i^{c}({\bf x},t){\bf
c}_i$. $\zeta_i$ are the coefficients resulting from the velocity space
discretization and $c_s=1/\sqrt{3}$ is the speed of sound, both of which are
determined by the choice of the lattice. The kinematic viscosity of the fluid
is given by $\nu^c=c_s^2(\tau^c-1/2)$. 

The interaction between fluid components $c$ and $c'$ is introduced as a
self-consistently generated mean field force 
\begin{equation}
\label{eqn_shan-chen-F}
{\bf F}^c({\bf x},t) \equiv -\Psi^c({\bf x},t)\sum_{c'}g_{c c'}\sum_{\bf x^{\prime}}\Psi^{c'}({\bf x^{\prime}},t)({\bf x^{\prime}}-{\bf x})\mbox{ ,}
\end{equation}
where $\bf x^{\prime}$ are the nearest neighbors and
$\Psi^{c}\left(\mathbf{x}\right)$ is the so-called effective mass, which can
have a general form for modeling various types of fluids. We choose
\begin{equation}
    \Psi^{c}\left(\mathbf{x},t\right)=\rho_0\left(1-\exp\left({-\frac{\rho^{c}\left(\mathbf{x},t\right)}{\rho_0}}\right)\right).
\label{eqn_ref_dens}
\end{equation}
$g_{cc'}$ is a force coupling constant whose magnitude controls the strength of
the interaction between components $c$, $c'$ and is set positive to mimic
repulsion.  The dynamical effect of the force is realized in the BGK collision
operator by adding to the velocity ${\bf u}$ in the equilibrium distribution
the increment
\begin{equation}
\Delta{\bf u}^{c}({\bf x},t) = \frac{\tau^{c}{\bf F}^{c}({\bf x},t)}{\rho^{c}({\bf x},t)}\mbox{ .}
\end{equation}
The force also enters the calculation of the actual macroscopic bulk
velocity
as~\cite{bib:guo-zheng-shi:2002,bib:jens-narvaez-zauner-raischel-hilfer:2010}
\begin{equation}
{\bf u}^{c}({\bf x},t) = \frac{ \sum_i  f_i^c \left( {\bf x},t \right) {\bf c}_i }{ \rho^c  \left( {\bf x},t \right)}  + \frac{1}{2} {\bf F}^{c}({\bf x},t).
\end{equation}

In this paper two fluids with identical properties are used which are called
``blue'' (``b'') and ``red'' (``r'').  To simplify statements about the fluid
ratio at a certain position an order parameter 
\begin{equation}
\phi\left(\mathbf{x},t\right)=\rho^\textnormal{r}\left(\mathbf{x},t\right) - \rho^\textnormal{b}\left(\mathbf{x},t\right)
\end{equation}
is introduced. The Shan-Chen model is a diffuse interface method, where
interfaces between different fluids are about four lattice sites wide. For the
analysis below we define the interface position to be located where the order
parameter vanishes.

\section{Suspended particles}
\label{sec_suspensions}
Pioneering work on the development of an extension to the lattice Boltzmann
method to incorporate particles has been done by Ladd et
al.~\cite{Ladd94a,Ladd94b,Ladd01}. The method has been applied to suspensions
of spherical and non-spherical particles by various
authors~\cite{bib:jens-herrmann-bennaim:2008,bib:aidun-lu-ding,bib:stratford-adhikari-pagonabarraga-desplat-cates-2005,bib:joshi-sun}.
Recently, the inclusion of Brownian motion was revisited and clarified in more detail~\cite{bib:adhikari-stratford-cates-wagner-05,bib:duenweg-schiller-ladd:2007}.
The suspended particles are assumed to be homogeneous spheres with radius
$r_\textnormal{par}$. In our implementation of the method Newton's equations
for the momentum
\begin{equation}
\mathbf{F}=m \cdot \frac{\mathrm{d} \mathbf{u}_\textnormal{par}}{\mathrm{d}t}
\end{equation}
and the angular momentum
\begin{equation}
\mathbf{D}=J \cdot \frac{\mathrm{d} \omega}{\mathrm{d}t}
\end{equation}
are solved with a leap frog integrator to simulate their behavior.
Here, $\mathbf{F}$ is the force acting on a particle, $m$ is its mass and
$\mathbf{u}_\textnormal{par}$ its velocity. $\mathbf{D}$ is the torque,
$J$ the moment of inertia and $\omega$ the angular velocity.

The particles are discretized on the lattice and interactions between the fluid
and the particles are introduced by marking all lattice sites that are inside
the particle as solid nodes for the fluid, at which bounce back boundary
conditions are applied~\cite{Ladd01}.  Bounce back boundary conditions reflect
the incoming distributions at a site back to where they came from, so that the
streaming step is modified to
\begin{equation}
f_i^c \left( \mathbf{x} , t + 1 \right) = f_i^c \left( \mathbf{x} - \mathbf{c}_i, t
\right),
\end{equation}
for all $i$ where $\mathbf{x} - \mathbf{c}_i$ is not a solid node to
\begin{equation}
f_i^c \left( \mathbf{x}, t + 1 \right) = f_{i'}^c \left( \mathbf{x}, t \right),
\label{eqn_bounce-back}
\end{equation}
for all $i$ where $\mathbf{x} - \mathbf{c}_i$ is a solid node. Here, $i'$ is
defined as the index corresponding to $\mathbf{c}_i = -\mathbf{c}_{i'}$. This
results in a no-slip boundary located halfway between the fluid and the solid
node.
The change of momentum of the fluid that is reflected at the boundary has to be
compensated by a momentum change of the particle itself as given by
\begin{equation}
\Delta \mathbf{p}(t) = 2 \cdot \rho^c(\mathbf{x},t) \mathbf{c}_i.
\end{equation}
As we assume the length of a time step $\mathrm{d}t$ to be $1$ this corresponds to a force
\begin{equation}
\mathbf{F}(t)=2 \cdot \rho^c(\mathbf{x},t) \mathbf{c}_i
\label{eqn_bounce-back-force}
\end{equation}
and a torque
\begin{equation}
\mathbf{D}(t) = \mathbf{F}(t) \times \mathbf{r}(t),
\end{equation}
on the particle. Here, $\mathbf{r}(t)$ is the vector pointing from the center
of the particle to the site of the reflection.  Since the particles are not
stationary but move over the lattice, the bounce back rule does not correctly
reproduce the velocity of the reflected fluid. It is therefore corrected as
\begin{equation}
f_i^c\left( \mathbf{x}, t + 1 \right) = f_{i'}^c \left( \mathbf{x}, t \right) - \frac{1}{6} \rho^c\left(\mathbf{x},t\right) \mathbf{u}_\textnormal{surf}\left(\mathbf{x},t\right) \cdot\mathbf{c}_{i'},
\end{equation}
where $\mathbf{u}_\textnormal{surf}(\mathbf{x},t)$ is the velocity of the
particle surface on which the fluid is reflected. This effect also leads to a
change in transfered momentum, so that the force acting on the particle is
\begin{equation}
\label{eqn_ladd_bewegtewand}
\mathbf{F}(t)=\left( 2 \rho^c(\mathbf{x},t)  - \frac{1}{6} \rho^c(\mathbf{x},t) \mathbf{u}_\textnormal{surf}(\mathbf{x},t)\cdot\mathbf{c}_{i'} \right) \mathbf{c}_{i'}.
\end{equation}
When the particle moves over the lattice, individual lattice sites can be
either occupied by it in front of the particle or be released at its back. In
case of newly occupied sites, the fluid on the site is deleted and its momentum
transfered to the particle by adding
\begin{equation}
\mathbf{F}(t)=- \rho^c(\mathbf{x},t) \mathbf{u}^c(\mathbf{x},t).
\end{equation}
In case of the particle vacating a lattice site new fluid is created with the
initial fluid density and the velocity of the particle surface
$\mathbf{u}_\textnormal{surf}(\mathbf{x},t)$ at the corresponding site:
\begin{equation}
\label{eq:newfi}
f_i^c(\mathbf{x},t)=\rho^c_\textnormal{init} \cdot f_i^{\textnormal{eq}^c} \left( \mathbf{u}_\textnormal{surf}(\mathbf{x},t),\rho(\mathbf{x},t) \right).
\end{equation}
To satisfy conservation of momentum this again leads to a force on the
particle, 
\begin{equation}
\label{eq:Fcorrsurf}
\mathbf{F}(t)= \rho^c_\textnormal{init} \mathbf{u}_\textnormal{surf}(\mathbf{x},t).
\end{equation}

Interactions between particles can be taken into account similar to standard
molecular dynamics implementations. In the current paper, we only consider
Hertzian contact forces to mimic hard spheres and a lubrication correction to
correct for the limitations of the lattice Boltzmann method to describe the
hydrodynamics properly on scales below the lattice resolution.
When particles collide the resulting forces are derived from the Hertzian potential~\cite{bib:hertz}
\begin{equation}
\label{eq:hertz}
V_{\rm Hertz} \left( r \right)=
\begin{cases}
K_\textnormal{Hertz} \cdot \left(2 r_\textnormal{par} - r
\right)^{\frac{5}{2}}  & \mbox{for } r < 2 r_\textnormal{par}\\
0 & \mbox{else,} \\
\end{cases}
\end{equation}
with
$K_\textnormal{Hertz}$ being a
constant~\cite{bib:jens-hecht-ihle-herrmann:2005}.  When two particles move
towards each other the lubrication interaction between them results in a force
separating the particles. If there is not at least one lattice site between the
particles to resolve the flow, this force is not properly reproduced by the
simulation and a lubrication correction has to be added as given by
\begin{equation}
\label{eqn_ladd_lubrikation}
\mathbf{F}_\textnormal{lub} = - \frac{6 \pi \nu^c r_\textnormal{par}^4}{\left(2 r_\textnormal{par}\right)^2} \frac{\mathbf{r}}{\left|\mathbf{r}\right|} \left[ \frac{\mathbf{r}}{\left|\mathbf{r}\right|} \left( \mathbf{u}_1 - \mathbf{u}_2 \right)\right] \left( \frac{1}{\left|\mathbf{r}\right| - 2r_\textnormal{par}}  - 1    \right),
\end{equation}
where $\mathbf{r}$ is the vector connecting the particle centers and
$\mathbf{u}_{1,2}$ is their respective velocity~\cite{Ladd01,Hecht07}.  It
is introduced at particle distances smaller than $\frac{2}{3}$ lattice
units and limited to it's value at a distance of  $\frac{1}{10}$ of a
lattice unit to avoid numerical instabilities due to the divergence at
$\left|\mathbf{r}\right| = 2r_\textnormal{par}$.

\section{Particles in multicomponent fluids}
\label{sec_combination}
In order to develop a simulation algorithm for particles in multicomponent
flows the previously described methods can be combined as described in the
current section. 
First, when extending the coupling between particles and fluid to multiple
components the treatment of lattice sites that are uncovered by the moving
particle has to be adapted. In the original algorithm for a single fluid, such
sites are filled with the initial fluid density $\rho^c_{\rm init}$. However,
re-initializing such sites with multiple fluid components can lead to
artefacts since it is not a priori the case that the correct fluid
composition should correspond to the initial state of the simulation. For
example, one kind of fluid could appear in a region where only the other fluid
is present. To prevent such artefacts we use the average surrounding fluid
density
\begin{equation}
\bar{\rho}^c(\mathbf{x},t) =\frac{1}{N_\textnormal{NP}} \sum_{i_\textnormal{NP}} \rho^c(\mathbf{x}+\mathbf{c}_{i_\textnormal{NP}},t),
\end{equation}
where $i_\textnormal{NP}$ are all indices $i$ for which
$\mathbf{x}+\mathbf{c}_i$ is a non-particle site. $N_\textnormal{NP}$ is
the number of these sites. Similar to the method described in
\cite{bib:aidun-lu-ding,bib:lorenz-caiazzo-hoekstra}, the uncovered sites
are re-initialized with 
\begin{equation}
\label{eqn_rho_neu}
\rho_\textnormal{new}^c(\mathbf{x},t)=\bar{\rho}^c(\mathbf{x},t)
\end{equation}
following a similar approach as in Eqs.~\ref{eq:newfi} and
\ref{eq:Fcorrsurf}, i.e.
\begin{equation}
f_i^c(\mathbf{x},t)=\rho_\textnormal{new}^c(\mathbf{x},t) \cdot f_i^\textnormal{eq} \left(\rho_\textnormal{new}^c(\mathbf{x},t), \mathbf{u}_\textnormal{surf}(t) \right)
\end{equation}
and
\begin{equation}
\label{eqn_ladd_vernichtenunderzeugen}
\mathbf{F}=\sum_c \rho_\textnormal{new}^c(\mathbf{x},t) \mathbf{u}_\textnormal{surf}(t).
\end{equation}

The second modification of the original algorithms is required to correctly
take into account the effect of the fluid-fluid interaction forces on the fluid
in the direct vicinity of a particle.  For the calculation of the forces
between different fluid components, also the empty lattice sites inside a
particle are considered if one follows Eq.~\ref{eqn_shan-chen-F}.  Since there
are no Shan-Chen forces acting in the direction from the particle to the fluid,
the fluid forms a layer of increased density around the particle. To avoid this
artefact, the outermost layer of lattice sites inside the particle is not kept
empty, but is filled with a virtual fluid density which is equivalent to the
average of the surrounding densities $\bar{\rho}$:
\begin{equation}
  \rho^c(\mathbf{x},t) = \bar{\rho}^c(\mathbf{x},t)
\end{equation}
This virtual fluid inside the particles does not follow the lattice Boltzmann
equation, i.e. the advection and collision steps are not applied. 

\begin{figure}
\includegraphics[width=1.0\linewidth]{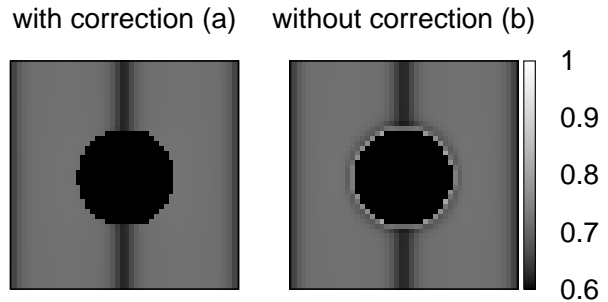}
\caption{$\sum_c \rho^c$ after 2000 timesteps in the presence of a particle with
$r_\textnormal{par}=10$ at an interface at the center of the shown area.
The cut on the left (a) shows a particle being filled with virtual fluid,
while on the right (b) the particle is empty as in the original algorithm.
Without the virtual fluid, the halo of increased density can clearly be
seen, while adding the virtual fluid successfully allows to correct for
this inconsistency. The parameters of the simulation were $\rho_0 = 1$ and
$\tau = 1$ for both species, the system of size $48^3$ lattice sites was
initially divided into two lamellae of width 24 lattice sites with density
$\rho^\textnormal{r}=\rho^\textnormal{b}=0.7$, respectively. All units are
given in lattice units throughout this paper.}
\label{fig_methode-totaldensity}
\end{figure}

Further, the Shan-Chen like force acting from the fluid surrounding a
particle on the particle itself has to be accounted for. This is implemented by
summing up all Shan-Chen forces
\begin{equation}
   \label{eqn_shan_chen_particle}
{\bf F}(t) = \sum_{{\bf x}} \sum_{c} {\bf F}^c({\bf x},t) 
\end{equation}
acting on every lattice site ${\bf x}$ inside the particle.
This force and the corresponding torque are then added to the particle within
the molecular dynamics algorithm. The forces on the fluid outside the
particles are calculated as before with the virtual fluid being treated
like a regular fluid in the Shan-Chen force computation. This leads to a
balanced force on the fluid sites near the particle surface and therefore
prevents the formation of a layer of increased density. 
This is demonstrated in Fig.~\ref{fig_methode-totaldensity}, where the
left subfigure shows a particle with $r_\textnormal{par}=10$ being filled
with virtual fluid while in the right subfigure the particle is not filled
with a virtual fluid. The particle is set at an interface created by two
lamellae of red and blue fluids at the center of
the shown area. Periodic boundary conditions are applied causing a second
interface to appear at the left and right borders of the sketches.
As we use a diffuse interface method for the fluids, the interfaces cover
about four lattice sites depicted by the varying grey scale.
Without the virtual fluid, the halo of increased density can clearly be 
seen, while adding the virtual fluid successfully allows to correct for 
this inconsistency. 

The advantage of a virtual fluid inside the particles is that it can be
utilized to modify the wettability of the particles. 
Here, we follow an approach which has been introduced to model hydrophobic
fluid-surface interactions for studying flow in hydrophobic microchannels
or droplets on surfaces with arbitrary contact
angles~\cite{bib:jens-kunert-herrmann:2005,bib:jens-kunert:2008c,bib:jens-jari:2008,bib:jens-schmieschek:2010}.
Our approach is equivalent to the method presented
in~\cite{bib:onishi-kawasaki-chen-ohashi}.  The Shan-Chen interaction between
the particles and the fluids can be
modified by tuning the density of the local virtual fluids. Increasing
one of them by an amount $\left|\Delta\rho\right|$ causes the particle
surface to ``prefer'' this fluid with respect to the other one, i.e. the
repulsion between the increased component and the unmodified one increases.
$\Delta\rho$ is called ``particle color'' and a positive particle color is
defined as an addition of ``red'' fluid, i.e.
\begin{equation}
\rho_{\rm new}^{\rm r}=\bar{\rho}^{\rm
r} + \left| \Delta\rho \right|,
\end{equation} 
whereas a negative color corresponds to ``blue'' fluid being added, i.e.
\begin{equation}
\rho_{\rm new}^{\rm b}=\bar{\rho}^{\rm b} +
\left| \Delta\rho \right|.
\end{equation}
In the next section we demonstrate that the particle color can be used to
tune the contact angle of the particle surface at an interface in order to
resemble specific fluids and solid materials. As an alternative to the
virtual fluid, a modified version of
Eq.~\ref{eqn_shan-chen-F} that takes solid lattice sites and fluid-surface
interactions into account could be developed, but the approach presented
here is simpler to implement and does not have any relevant impact on the
performance of the code.

The changing discretization of the particle together with the
fluid-surface interaction force leads to slight mass errors during the
step in which vacated lattice nodes are refilled with fluid. This effect
is especially strong for small particles and high forces (large values for
$g_\textnormal{cc'}$). Typical test cases have shown that the total mass
after very long simulation times ($10^7$ timesteps) increases by about one
percent. This can be explained by the simple interpolation for the amount
of newly
created fluid which is necessary since no analytical solution for the
multicomponent Shan-Chen model is known that describes the density profile
at interfaces. Even though the effect is very small, it can be suppressed
if newly created fluid densities $\rho_\textnormal{new}^c$ are scaled with
a correction factor which depends on the total mass error $\Delta \rho^c$
up to the current timestep and the total number of lattice sites in the
system $N$. This leads to a modification of Eq.~\ref{eqn_rho_neu}:
\begin{equation}
\label{masscorr}
\rho_\textnormal{new}^c=\bar{\rho}^c\left( 1 - C_0 \frac{\sum_c
\rho_\textnormal{init}^c}{\rho_\textnormal{init}^c} \frac{\Delta
\rho^c}{N}\right).
\end{equation}
The rate of the
adaptive correction can be tuned with the parameter $C_0$. Due to the very
small mass error, the correction can act very slowly, but should not be
chosen too fast in order to avoid hysteresis effects.
Further, Eq.~\ref{masscorr} reduces unphysical density gradients at
particle surfaces and thus contributes to the stability of the algorithm.
Repeating the same test case as above with a correction factor of $C_0=10$
results in a deviation of the mass of 0.03 percent after $10^7$ timesteps.

\section{Contact angle measurements}
\label{sec_contactangle}
The contact angle $\theta$ is a common measure for the wettability of the
particle by the two fluid components. The influence of various simulation
parameters on the contact angle is investigated in the current section.
\begin{figure}[h]
\centerline{
\includegraphics[width=0.8\linewidth]{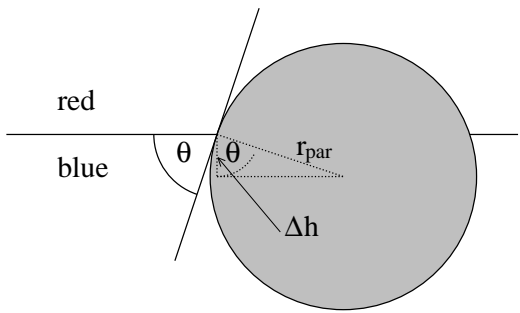}}
\caption{\label{fig_contactangle_def}
Definition of the contact angle $\theta$ for a particle of radius
$r_\textnormal{par}$ at the interface between the two fluid components.
$\Delta h$ is the distance from the particle center to the interface.
}
\end{figure}
In order to measure the contact angle the following setup is used in the
simulations: in $z$ direction, one half of the lattice is filled with one fluid
component, the other half with the second one. The particle is placed at the
interface at $t=0$ and the simulation is started. The interface between the two
components is tracked via the (linearly interpolated) position at which the
order parameter is zero. Then, the contact angle $\theta$ can be calculated by
\begin{equation}
\cos\left(\theta\right)=\frac{\Delta h}{r_\textnormal{par}},
\label{eqn_contactangle_calculation}
\end{equation}
where $\Delta h$ is the difference between the interface position and the
particle center in the direction perpendicular to the interface (cf. figure
\ref{fig_contactangle_def}).

\begin{figure}[h]
\centerline{
\includegraphics[width=1\linewidth]{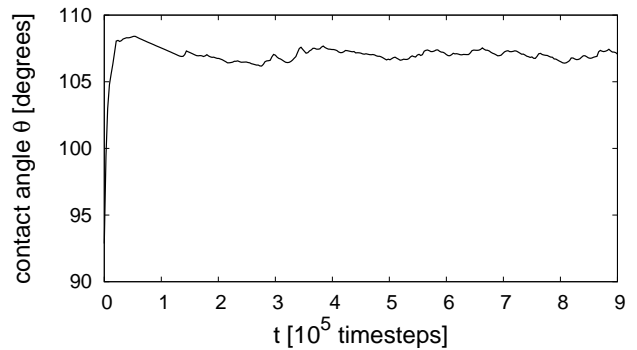}}
\caption{\label{fig_contactangle_t}Contact angle versus time for a 48x48x256
system with particle color 0.02, $r_\textnormal{par}= 10$,
$g_\textnormal{br}=0.08$, and fluid density $\rho_\textnormal{init}=0.6$.
After a rapid change from about $93^{\circ}$ to about $107^{\circ}$ the contact
angle stays at a constant value and only shows small oscillations with an
amplitude of about $1^{\circ}$.}
\end{figure}
To study the time-evolution of the contact angle a system of size 48x48x256
lattice sites is used. The particle has a radius of 10 lattice sites and a
color of 0.02. The initial fluid density is 0.6 and $g_\textnormal{br}=0.08$.
Figure~\ref{fig_contactangle_t} shows the time dependence of the contact angle.
After a short, fast movement at the beginning of the simulation the contact
angle oscillates slightly around a fixed value. Here and for all further graphs
in this section, we average the contact angle over the timesteps from $6 \cdot
10^5$ to $9 \cdot 10^5$ leading to a final value of $\theta=\left(107.03 \pm
0.26\right)^{\circ}$. The error is given by the standard deviation of the data.
Relating the variation of the angle to a change of the position of the
interface on the lattice with regards to the particle center results in $\Delta
h = \left(-2.93 \pm 0.04\right)$ lattice units. The error in the position
measurement is very small with respect to the lattice resolution.

\begin{figure}[htb]
\centerline{
\includegraphics[width=0.5\linewidth]{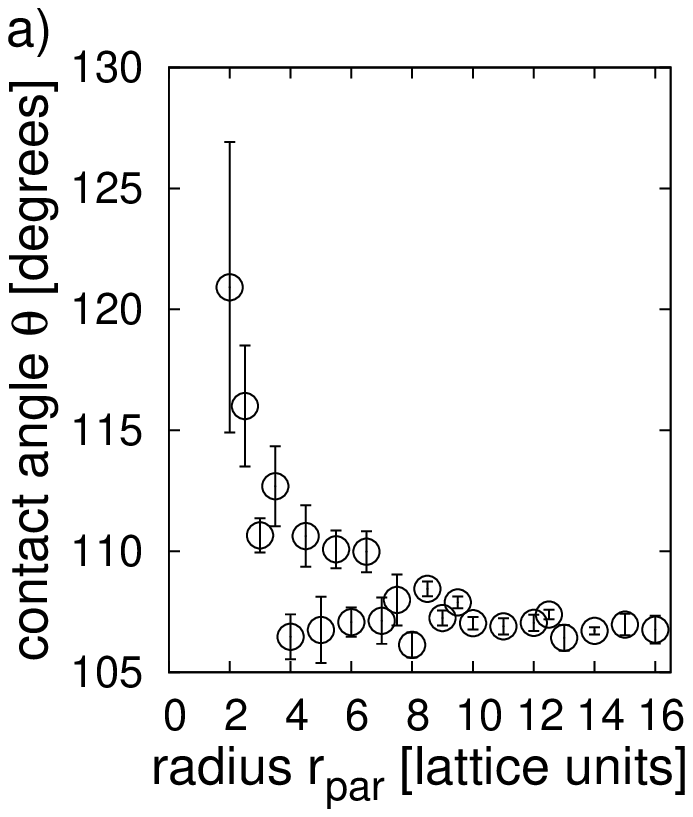}
\includegraphics[width=0.5\linewidth]{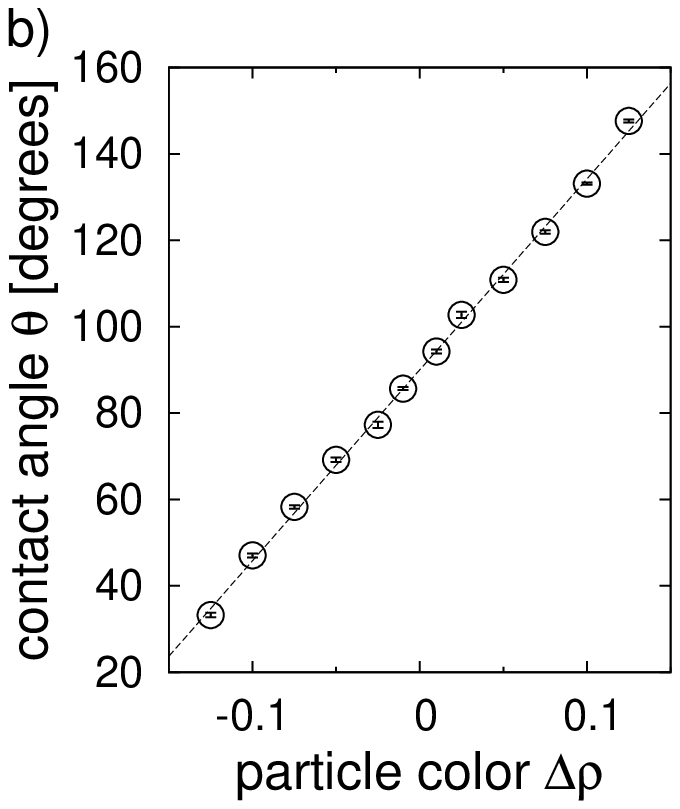}}
\caption{a) Contact angle versus particle size. The system size is 48x48x256, the
particle color 0.02, $g_\textnormal{br}=0.08$, and
$\rho_\textnormal{init}=0.6$. The average measured angle and errors given by
the standard deviation over timesteps from $6\cdot 10^5$ to $9 \cdot 10^5$ are
shown. The contact angle for $r_\textnormal{par}=2$ is $\left( 120.9 \pm 6.0
\right)^{\circ}$, for particles larger than $r_\textnormal{par}=5$ the error
stays below $1.5^{\circ}$ and the angles are in the range of $106^{\circ}$ to
$110^{\circ}$.  Particles with a non-integer radius show larger contact angles,
which can be adhered to a discretization effect.
\label{fig_kw_r}
\\
b) Contact angle versus particle color for $\rho_\textnormal{init}=0.7$. The
data can be fitted with the
equation $\theta=442 \cdot \Delta \rho + 90$ (dashed line).}
\label{fig_kw_f}
\end{figure}

Figure~\ref{fig_kw_r}a shows the resulting contact angle for different particle
sizes between $r_\textnormal{par}=2$ and $r_\textnormal{par}=16$. For small
particle radii the error increases substantially and the measured angle is not
equivalent to the one measured for larger particles, but is up to $15^{\circ}$
larger.  For example, for $r_\textnormal{par}=2$ the contact angle is $\left(
120.9 \pm 6.0 \right)^{\circ}$.  For particles larger than
$r_\textnormal{par}=5$ the error stays below $1.5^{\circ}$ and the measured
angles are in the range of $106$ to $110^{\circ}$.
Smaller particles are more susceptible to small forces because of their smaller
mass, also one has to keep in mind that our lattice Boltzmann multicomponent
model is a diffuse interface method. Since the interface is about four lattice
sites wide, small particles are completely inside the interface region. For
particles with a non-integer radius the error and the angle are larger than for
particles with integer radii. This can be adhered to being a discretisation
effect as the particles move on the lattice.

The dependency of the contact angle $\theta$ on the particle color $\Delta\rho$
is shown in Fig.~\ref{fig_kw_f}b. One can see an almost linear relation between
the contact angle and the particle color in the range from a color of
$\Delta\rho=-0.125$ (contact angle $33.2^\circ$) to $\Delta\rho=0.125$ (contact
angle $147.6^\circ$). Included in the figure is a linear fit given by
$\theta=442 \cdot \Delta \rho + 90$ to stress this linear behavior. 
The simulations with a particle color of $\Delta\rho\ge 0.15$ and
$\Delta\rho\le-0.15$ result in a detachment of the particle from the
interface.
Thus, it is possible to choose a specific particle color to obtain the related
contact angle or to force detachment from one of the fluids.

\begin{figure}[htb]
\centerline{
\includegraphics[width=0.5\linewidth]{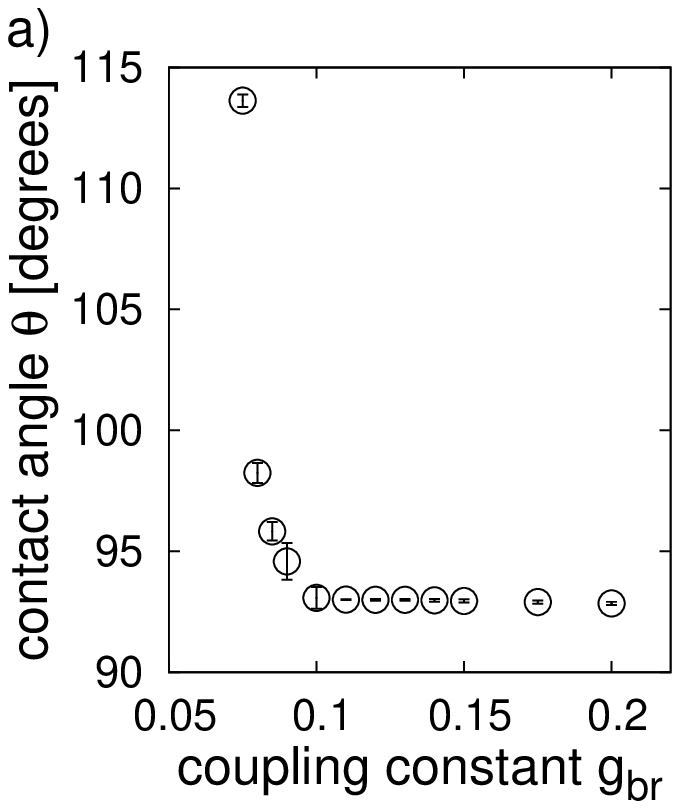}
\includegraphics[width=0.5\linewidth]{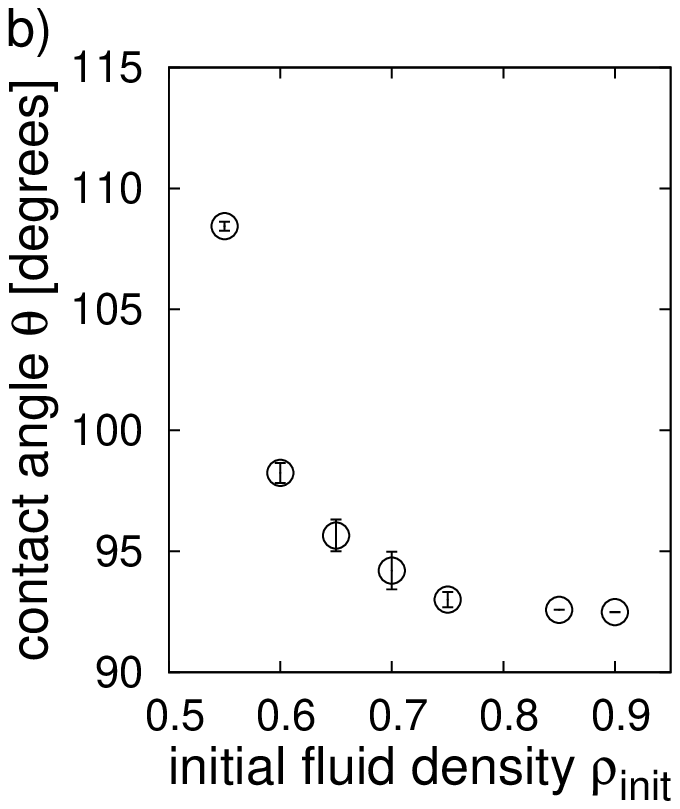}}
\caption{a) Contact angle versus $g_\textnormal{br}$. For $g_\textnormal{br}\ge
0.1$ a contact angle of $93.0^{\circ}$ with an error smaller than $0.1^{\circ}$
is measured. The contact angle increases from this value at
$g_\textnormal{br}=0.1$ to $98.0^{\circ}$ at $g_\textnormal{br}=0.08$ with an
error below one degree.  $g_\textnormal{br}=0.075$ results in a contact angle
of $\left(113.6 \pm 0.3 \right)^{\circ}$ and for smaller $g_\textnormal{br}$ the
particle detaches from the interface.\\
\label{fig_kw_g_br}
b) Contact angle versus the initial fluid density $\rho_\textnormal{init}$.
$\rho_\textnormal{init}=0.9$ results in a contact angle of $92.5^{\circ}$. 
Reducing $\rho_\textnormal{init}$ causes the contact angle to decrease to
$108.4^{\circ}$ at $\rho_\textnormal{init}=0.55$.}
\end{figure}

As a next step we investigate the influence of the strength of the fluid-fluid
interaction force on the contact angle. For strong forces determined by large
$g_\textnormal{br}$ the interface is well defined and the surface tension high.
Low $g_\textnormal{br}$ cause a low surface tension and thus a more diffuse
interface. When the coupling constant $g_\textnormal{br}$ is varied, the
contact angle $\theta$ changes as shown in Fig.~\ref{fig_kw_g_br}a. The
stronger the force the stronger the particle is kept at the interface. If the
force is too weak the particle cannot be held at the interface anymore. For
$g_\textnormal{br} \geq 0.1$ almost no change to the contact angle can be
observed and $\theta$ converges to $93.0^{\circ}$ with an error smaller than
$0.1^{\circ}$. For $g_\textnormal{br} \leq 0.08$ the contact angle increases
dramatically until the particle does not stay attached to the interface at
$g_\textnormal{br}\le 0.07$. On the one hand, a well defined contact angle and
well defined interfaces are preferrable requiring large values of
$g_\textnormal{br}$. On the other hand, too large values can cause very high
local flow velocities and the lattice Boltzmann method can become unstable.
Thus, $g_\textnormal{br}$ should be chosen as small as possible.

As can be observed in Fig.~\ref{fig_kw_g_br}b, a variation of the initial fluid
density $\rho_\textnormal{init}$ has a similar effect as a modification of the
coupling constant on the contact angle. However, $\theta$ only changes from
$108.5^{\circ}$ ($\rho_\textnormal{init}=0.55$) to $92.5^{\circ}$
($\rho_\textnormal{init}=0.9$), i.e. the effect is much weaker. The reason
for the lower impact on $\theta$ is given by our particular choice of the
effective mass $\Psi^{c}\left(\mathbf{x}\right)$ (see
Eq.~\ref{eqn_ref_dens}) which causes a damping of the interactions for
large densities.

The knowledge of the contact angle and particle shape together with a
measurement of the surface tension between both fluids allows to measure the
energy required to detach a trapped particle from an
interface~\cite{bib:binks-horozov}. While for spherical particles it is
straightforward to compute the detachment energy analytically, for highly
anisotropic or complex shaped particles this is not easily possible. A
simulation study based on the model proposed in this paper would allow a well
founded understanding of the dependence of detachment energies on particle
properties and could be compared to experimental data.

\section{Bijel formation}
\label{sec_bijel}
The formation of a ``bijel'' (bicontinuous interfacially jammed emulsion gel)
was first predicted by Stratford et. al. in
2005~\cite{bib:stratford-adhikari-pagonabarraga-desplat-cates-2005}. As stated
in the introduction, bijels can form when (colloidal) particles are added to a
mixture of two immiscible fluids. During the phase separation of the two
fluids, the particles accumulate at the interface until those are fully jammed.
Since the simulations performed by Stratford et. al. utilize a free energy based
multiphase lattice Boltzmann model, we show in this section that the
multicomponent model introduced in this paper is also able to model the
formation of a bijel. We study the temporal development of the system and
compare our results with the results of Stratford et al.. Further, we
investigate the influence of the particle concentration, $g_\textnormal{br}$,
and $\rho_\textnormal{init}=\sum_c \rho_\textnormal{init}^c$ on the bijel
formation. The initial conditions for the simulations are as follows: an
identical amount of the two fluid species is distributed randomly throughout
the system. The initial positions of the colorless particles
($\theta=90^\circ$) are also chosen at random. In order to keep the system size
at manageable $256^3$ lattice units and to be able to simulate a significant
number of particles, the particle radius is kept at $r_\textnormal{par}=5$
lattice units in all simulations. 

The conversion from lattice units to SI units of a system containing two
identical fluids with the speed of sound ($c_s=1482.35$m/s) and kinematic
visosity ($\nu=1.004\cdot10^{-6}$m$^2$/s) of water at 20$^\circ$C results in a
timestep of $\Delta t=9.14\cdot 10^{-13}$s and a lattice constant of $\Delta
x=2.35$nm. Since we set our particle diameter to 10 lattice units, the physical
diameter would be $23.5$nm and the side length of a cubic simulation volume with
$256$ lattice units corresponds to $601.6$nm. The systems presented in this
section are simulated for $2.8 \cdot 10^{5}$ timesteps corresponding to $2.56
\cdot 10^{-7} \textnormal{s}$.

\begin{figure}[htb]
\centerline{
\includegraphics[width=1\linewidth]{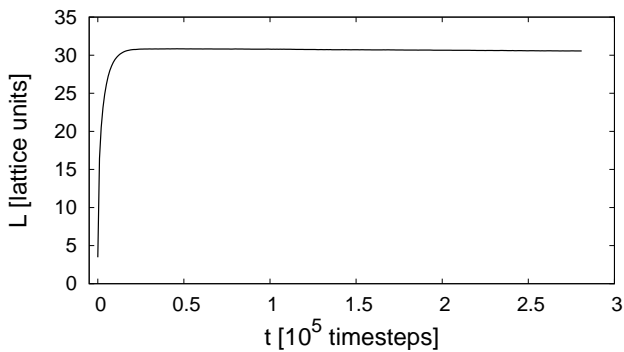}}
\caption{Average domain size for a system with $g_\textnormal{br}=0.08$,
$\rho_\textnormal{init}=0.7$ and a particle concentration of 20\%. During the
first $2.5 \cdot 10^4$ steps of the simulation the average domain size
grows from below 5 to its final value of about 31 lattice units.
}
\label{fig_jamming}
\end{figure}
As already stated correctly by Stratford et al., thermal fluctuations have only
little effect on the phase separation and bijel formation and can thus be
ignored in the
simulations~\cite{bib:stratford-adhikari-pagonabarraga-desplat-cates-2005}. 

To analyze the development of structures in the simulated systems, we
define the averaged time dependent lateral domain size $L(t)$ which
consists of an average of its components $L_i$ along direction $i=x,y,z$
as given by
\begin{equation}
\label{eq:domsize}
L_i(t)\equiv \frac{2\pi}{\sqrt{\left< k^2_i(t)\right>}}.
\end{equation}
Here,
\begin{equation}
\left<k^2_i(t)\right>\equiv \frac{\sum_\mathbf{k} k_i^2 S(\mathbf{k},t)}
{\sum_\mathbf{k} S(\mathbf{k},t)} 
\end{equation}
is the second order moment of the three-dimensional structure function
\begin{equation}
S(\mathbf{k},t)\equiv\frac{1}{N}\left|\phi^\prime_\mathbf{k}(t)\right|^2
\end{equation}
with respect to the Cartesian component $i$, $\left< \right>$ denotes the
average in Fourier space, weighted by $S(\mathbf{k}, t)$ and $N$ is the
number of nodes of the lattice, $\phi^\prime_\mathbf{k}(t)$ the Fourier
transform of the fluctuations of the order parameter
$\phi^\prime\equiv\phi-\left<\phi\right>$, and $k_i$ is the $i$th
component of the wave vector~\cite{bib:jens-giupponi-coveney:2007}.
The simulations are performed using a $256^3$ lattice, a coupling constant of
$g_\textnormal{br}=0.08$, an initial fluid density of
$\rho_\textnormal{init}=0.7$, a particle volume ratio $\alpha$ of 20 percent (about 6400
particles), a particle size of $r_\textnormal{par}=5$ lattice units, and a
particle density of 1, i.e. the particles are slightly heavier than the fluid.
The time development of the average domain size $L(t)$ is shown in
Fig.~\ref{fig_jamming}.
The figure clearly shows that
the system comes to arrest after a brief period of phase separation.
During the first $2.5 \cdot 10^4$ timesteps the average domain size $L(t)$
increases from 5 lattice units to about 31 lattice units and stays at that
value until the end of the simulation at $t=2.8 \cdot 10^5$ timesteps.  This
qualitatively agrees to the results obtained by Stratford et
al.~\cite{bib:stratford-adhikari-pagonabarraga-desplat-cates-2005}.
However, their simulation does not converge to a fixed domain size, which
might be caused by the thermal motion incorporated in their model. While
thermal fluctuations are not strong enough to detach the particles from
the interface they might cause some local reordering of the particles and
therefore support further domain growth. This effect would be favored by
the small particle diameter used in the simulations presented
in~\cite{bib:stratford-adhikari-pagonabarraga-desplat-cates-2005} since
the particle size is of the same order or even smaller than the interface
thickness.

\begin{figure}[htb]
\includegraphics{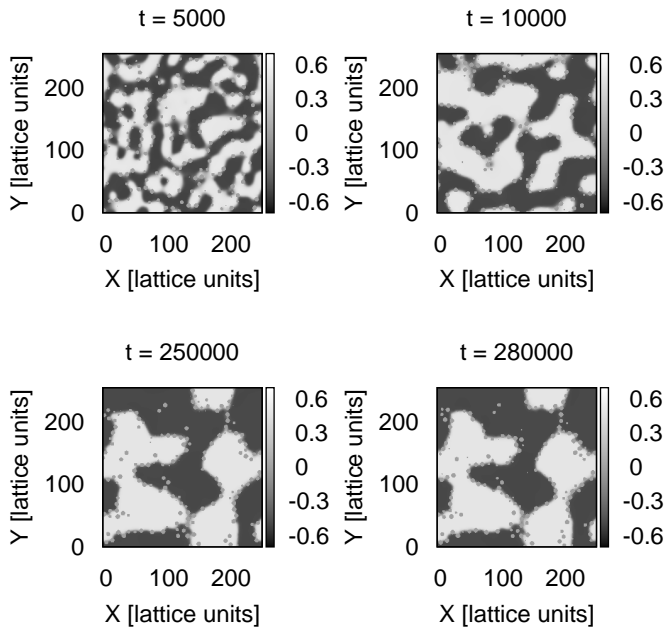}
\caption{2D cut of the order parameter $\phi$ at $z=0$ through the system studied in Fig.~\ref{fig_jamming}.
The spots where
$\phi=0$ correspond to the regions occupied by particles. $\phi$ shows
pronounced differences between $t=5000$ and $t=10000$, while at timesteps
$2.5\cdot  10^5$ and $2.8 \cdot  10^5$ only minor rearrangements can be
observed.}
\label{fig_jamming_colourfield}
\end{figure}
\begin{figure}[htb]
\begin{tabular}{cc}
\includegraphics[width=0.23\textwidth]{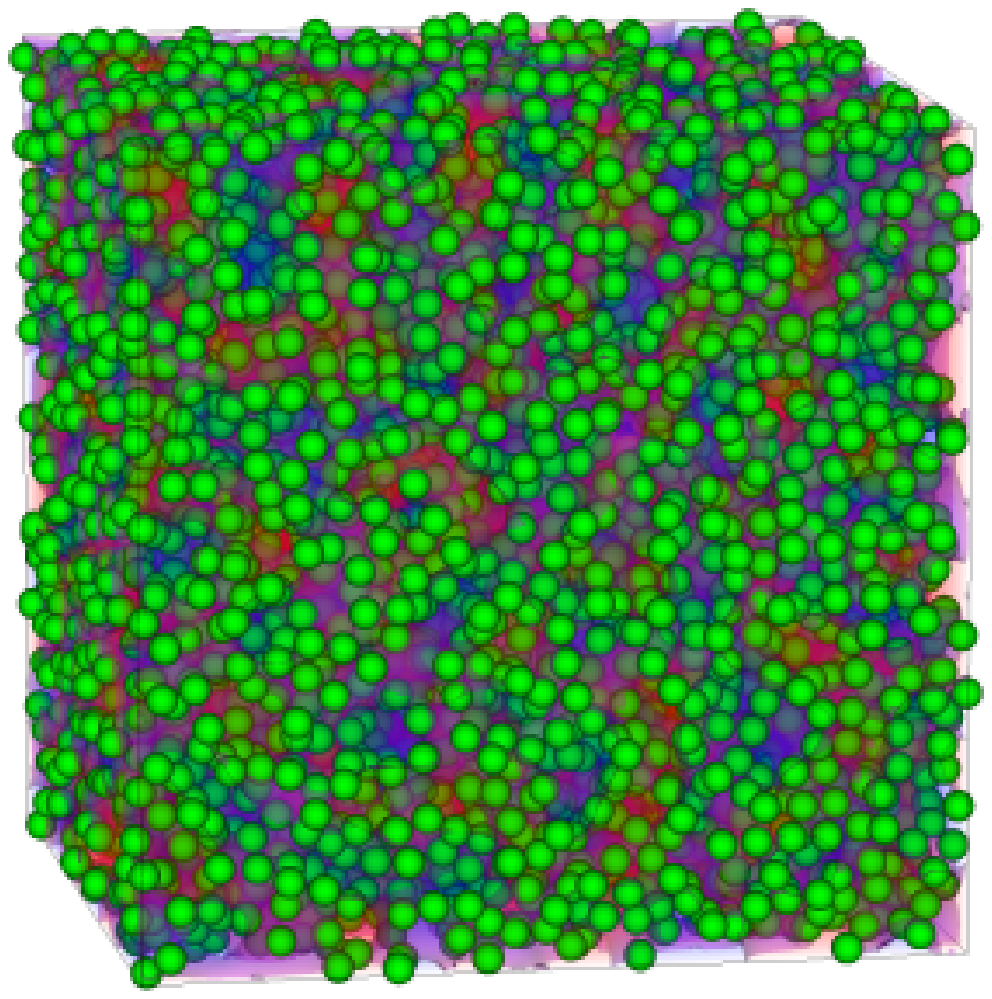} &
\includegraphics[width=0.23\textwidth]{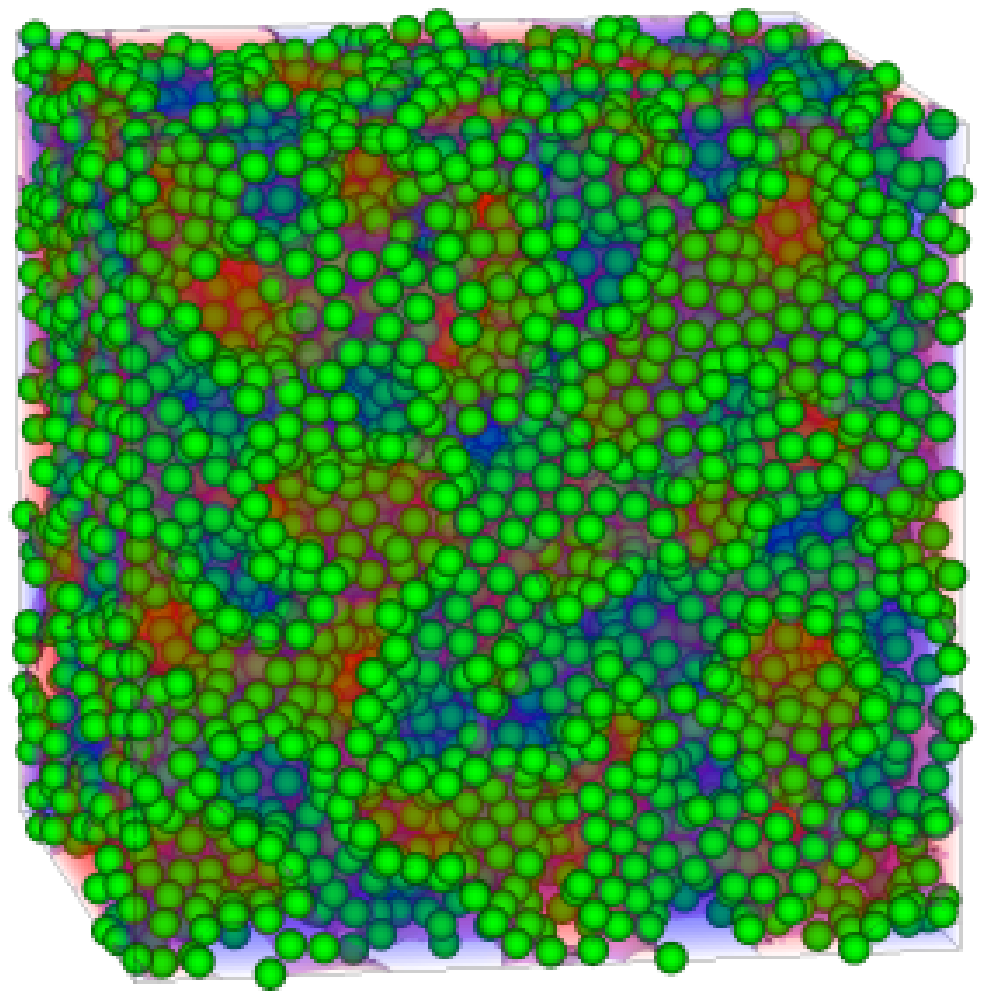} \\
$t=5000$ & $t=10000$ \\
\includegraphics[width=0.23\textwidth]{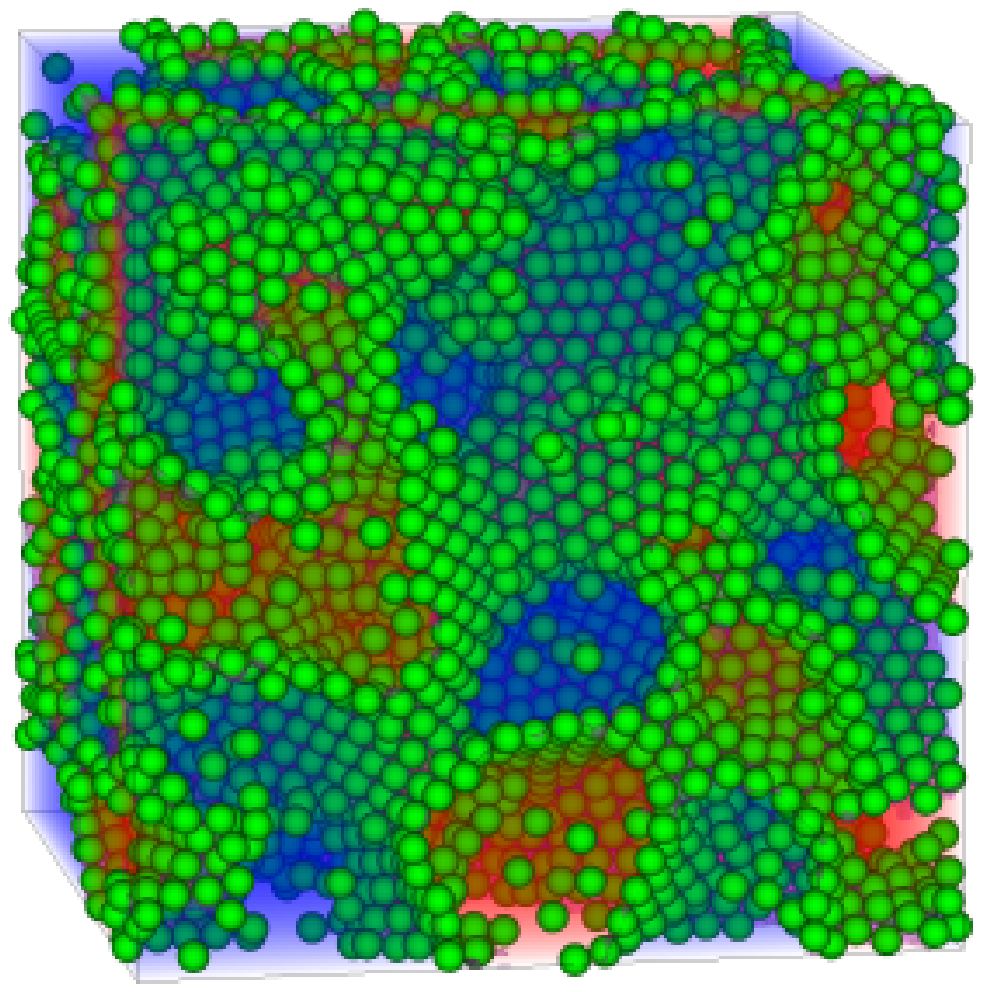} &
\includegraphics[width=0.23\textwidth]{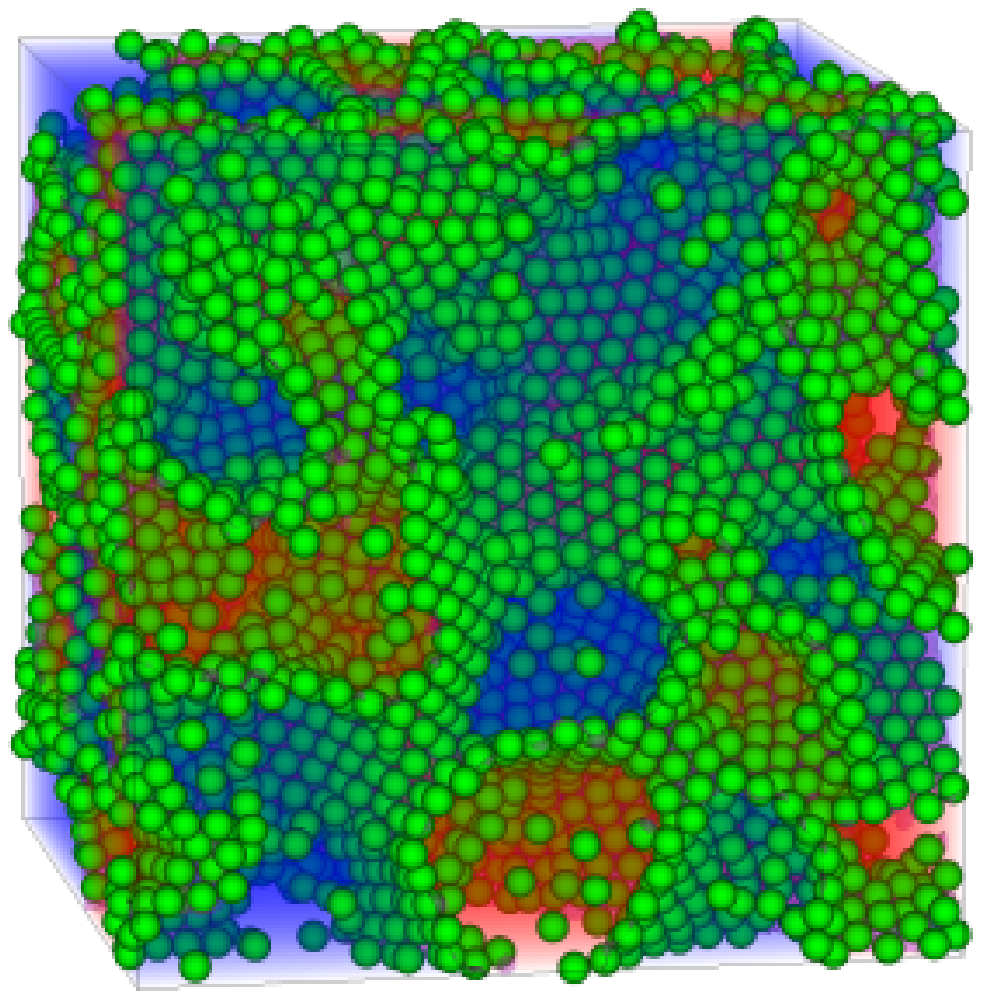}\\
$t=250000$ & $t=280000$ \\
\end{tabular}
\caption{(Color online) 3D visualisation of the system presented in
Figs.~\ref{fig_jamming} and \ref{fig_jamming_colourfield}. Shown are the
particles (in green/light gray) and the two fluids (in red/medium gray and blue/dark gray, respectively). The
visualizations for $t=2.5 \cdot 10^5$ and $t=2.8 \cdot 10^5$ nicely depict the
bicontinuouity of the fluids and the attachment of the particles to the
interface.}
\label{fig_jamming_images}
\end{figure}

The arrest of the phase separation process can also be observed by visualizing
a 2D cut at $z=0$ of the order parameter as in
Fig.~\ref{fig_jamming_colourfield} or a 3D visualisation of $\phi$ as in
Fig.~\ref{fig_jamming_images}.  The differences between timesteps $t=5000$ and
$t=10000$ are large while the system barely changes between timesteps $2.5
\cdot 10^5$ and $2.8 \cdot 10^5$. Both visualisations clearly demonstrate the
bicontinuouity of the fluid domains. In particular
Fig.~\ref{fig_jamming_images} depicts how the particles get trapped at the
fluid-fluid interface and cause the demixing process to stop.

\begin{figure}[htb]
\centerline{
\includegraphics[width=0.5\linewidth]{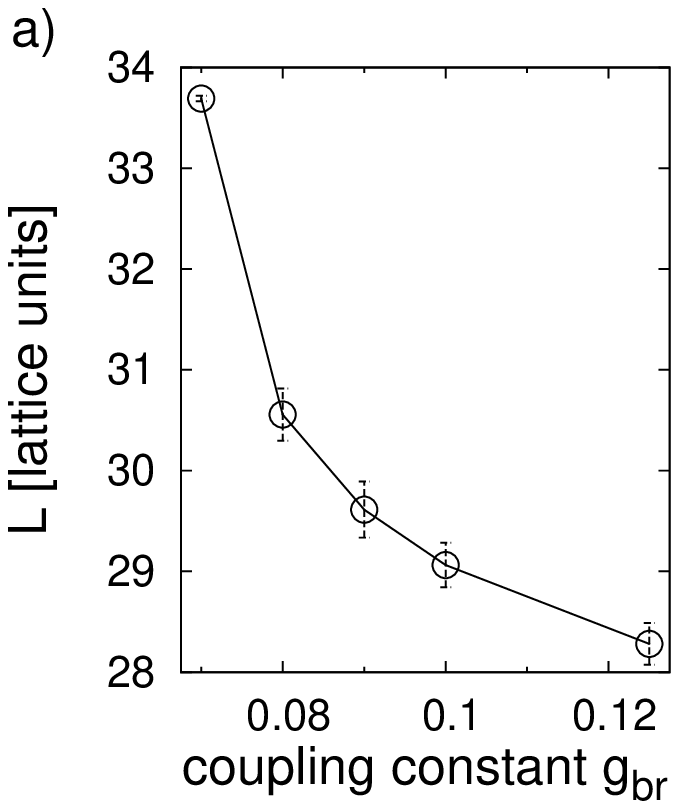}
\includegraphics[width=0.5\linewidth]{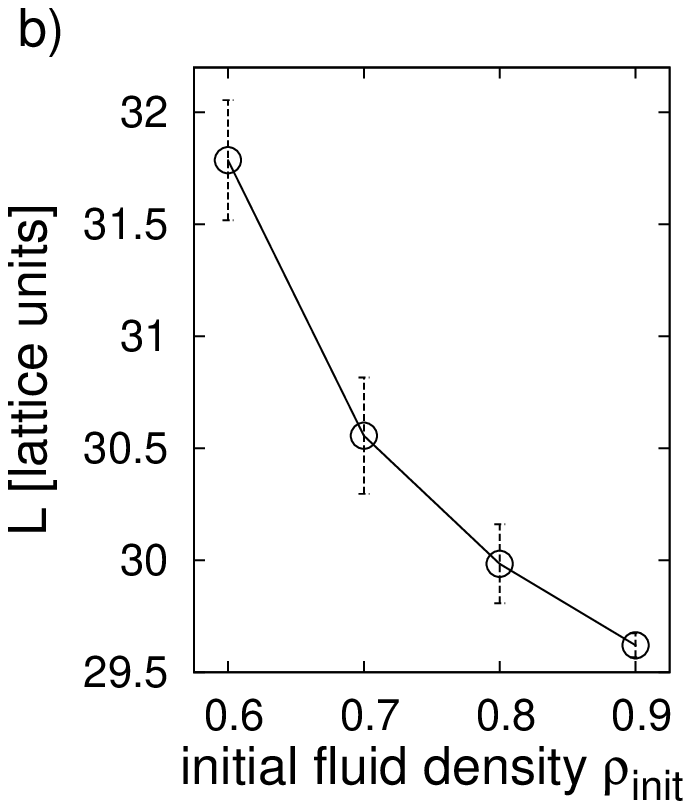}}
\caption{a) Average domain size versus $g_\textnormal{br}$ at $t=2.8\cdot
10^5$. With increasing $g_\textnormal{br}$ $L$ decreases from 33.7 lattice
units at $g_\textnormal{br}=0.07$ to 28 lattice units at
$g_\textnormal{br}=0.125$.  \\ \label{fig_jamming_gbr}
b) Average domain size versus $\rho_\textnormal{init}$.
$\rho_\textnormal{init}=0.6$ leads to a $L$=32 lattice units.
A larger value of $\rho_\textnormal{init}=0.9$ reduces $L$
to a value slightly below 30 lattice units. Error bars are given by the
maximum deviation of $L_x$, $L_y$, $L_z$ from the mean.}
\end{figure}

Modifying the strength of the fluid-fluid interaction force by varying the
coupling constant $g_\textnormal{br}$ also influences the resulting domain
size. This is demonstrated in Fig.~\ref{fig_jamming_gbr}a, where the averaged
lateral domain size $L$ is shown after $t=2.8 \cdot 10^5$ timesteps and for
different $g_\textnormal{br}$. While $g_\textnormal{br}=0.07$ leads to a domain
size of about 33.7 lattice units, $g_\textnormal{br}=0.125$ results in an
average size of 28 lattice units. The differences between the spatial
directions at a certain $g_\textnormal{br}$ are below 0.5 lattice units. A
higher value of the coupling constant leads to stronger forces attaching the
particles to the interface. Therefore, the size of the interface increases
because the particles cannot slightly shift away from it in order to accomodate
more particles on the same interfacial area. As the number of particles in the
system is kept constant, the interfacial area has to increase and therefore the
resulting domains become smaller.

For a variation of the initial bulk density $\rho_\textnormal{init}$
the arguments of the previous paragraph still hold. A larger value of
$\rho_\textnormal{init}$ leads to stronger interaction forces and
therefore to smaller structures as described above. While the coupling
constant $g_\textnormal{br}$ directly changes the strength of the force
$\rho_\textnormal{init}$ affects the force only indirectly through the
effective mass which causes the effect to be less pronounced.  As depicted
in Fig.~\ref{fig_jamming_gbr}b the average domain size $L$ decreases from
about 32 lattice units for $\rho_\textnormal{init}=0.6$ to below 30
lattice units for $\rho_\textnormal{init}=0.9$. 

\begin{figure}[htb]
\centerline{
\includegraphics[width=1\linewidth]{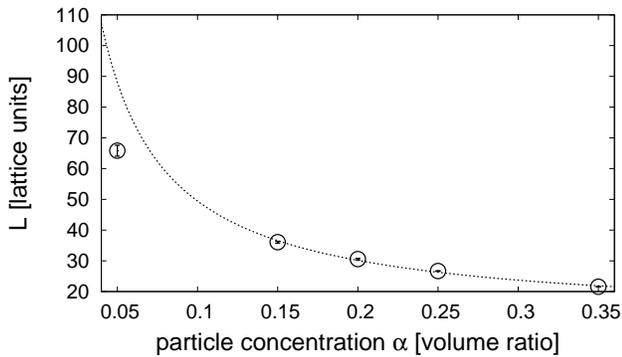}}
\caption{Average domain size versus particle concentration $\alpha$.
Decreasing $\alpha$ from 0.35 to 0.15 leads to an increase of $L$
from 21.5 lattice units to 36 lattice units. If the
concentration is further reduced to 0.05, finite size effects start to occur. Also shown is
$L=\frac{3.86}{\alpha}+10.85$, the result of a fit to the
concentration values between 0.15 and 0.35.
Error bars are given by the 
maximum deviation of $L_x$, $L_y$, $L_z$ from the mean.
}
\label{fig_jamming_c}
\end{figure}

The connection between the area of the interface covered by particles and the
size of the resulting structures can be best shown by varying the particle
concentration $\alpha$. This is depicted in Fig.~\ref{fig_jamming_c}.
Increasing particle concentration leads to a larger interfacial area and
therefore to finer structures. While the average domain size of a system with a
particle concentration of 0.15 is about 36 lattice units this value decreases
to about 22 lattice units for a concentration of 0.35.  A too low particle
concentration leads to such a small stabilized surface that finite size effects
start to appear as they are well known from lattice Boltzmann simulations of
spinodal decomposition~\cite{bib:jens-giupponi-coveney:2007}. This can be seen
for a concentration of 5\%. Here, the structure size increases drastically,
also the average domain size is not the same for all spatial directions anymore
and varies by about 3 lattice units. It is possible to fit a function of the
form $L={a}/{\alpha+b}$ with $a=3.85936$ and $b=10.8479$ to the values where
finite size effect do not play a role.

\section{Pickering emulsions}
\label{sec_emulsion}
\begin{figure}[htb]
\centerline{
\includegraphics[width=1\linewidth]{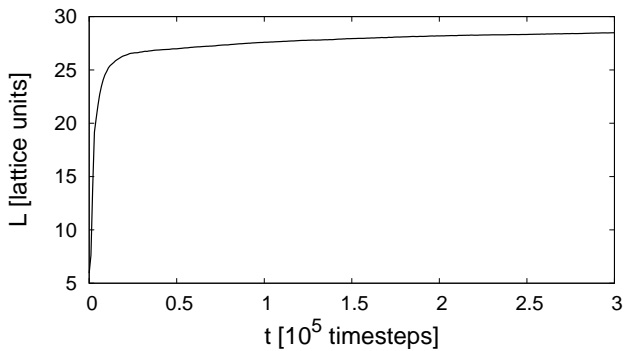}}
\caption{Average domain size over time for a system of size $256^3$,
$g_\textnormal{br}=0.09$, $\rho_\textnormal{init}=0.66$, fluid ratio 1:3,
particle color $-0.01$ and particle concentration $\alpha=0.15$. After a rapid
growth of the domain size to over 25 lattice units during the first 25000
timesteps the domain size increases only slowly to slightly over 27
lattice units at timestep $3.0 \cdot 10^5$.}
\label{fig_pickering_c0.15-L}
\end{figure}
\begin{figure}[htb]
\begin{tabular}{cc}
\includegraphics[width=0.23\textwidth]{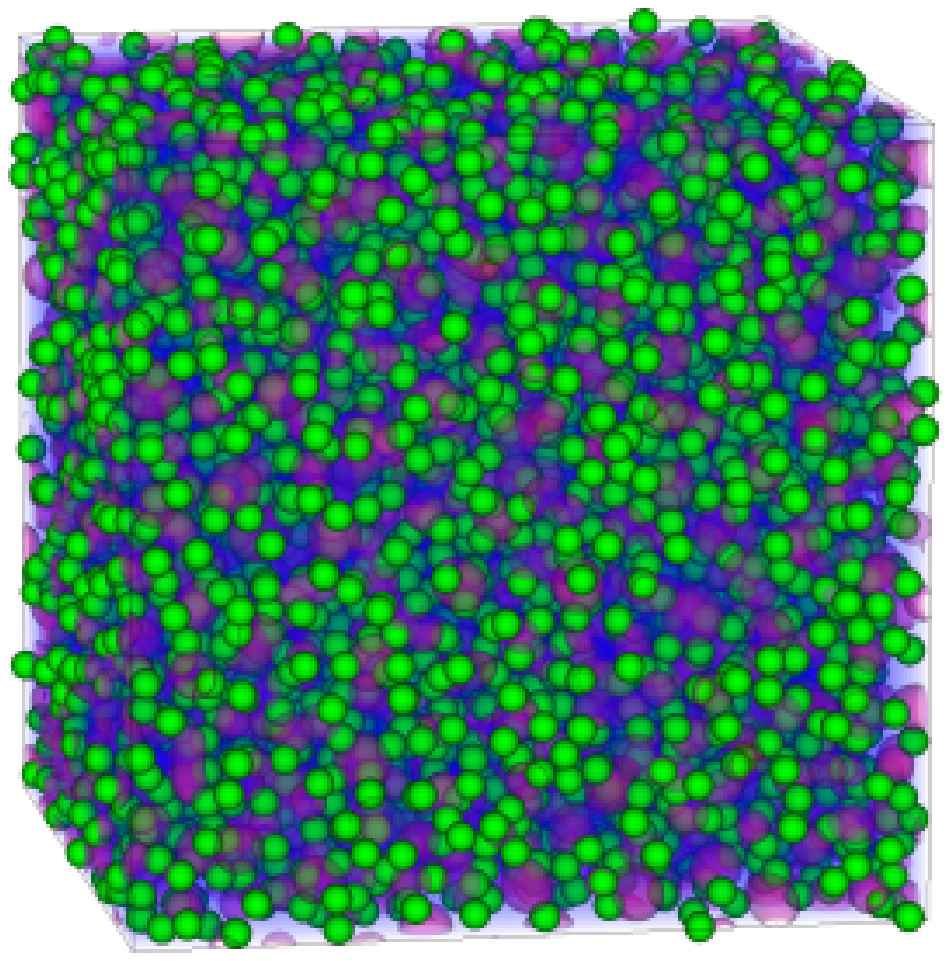} & 
\includegraphics[width=0.23\textwidth]{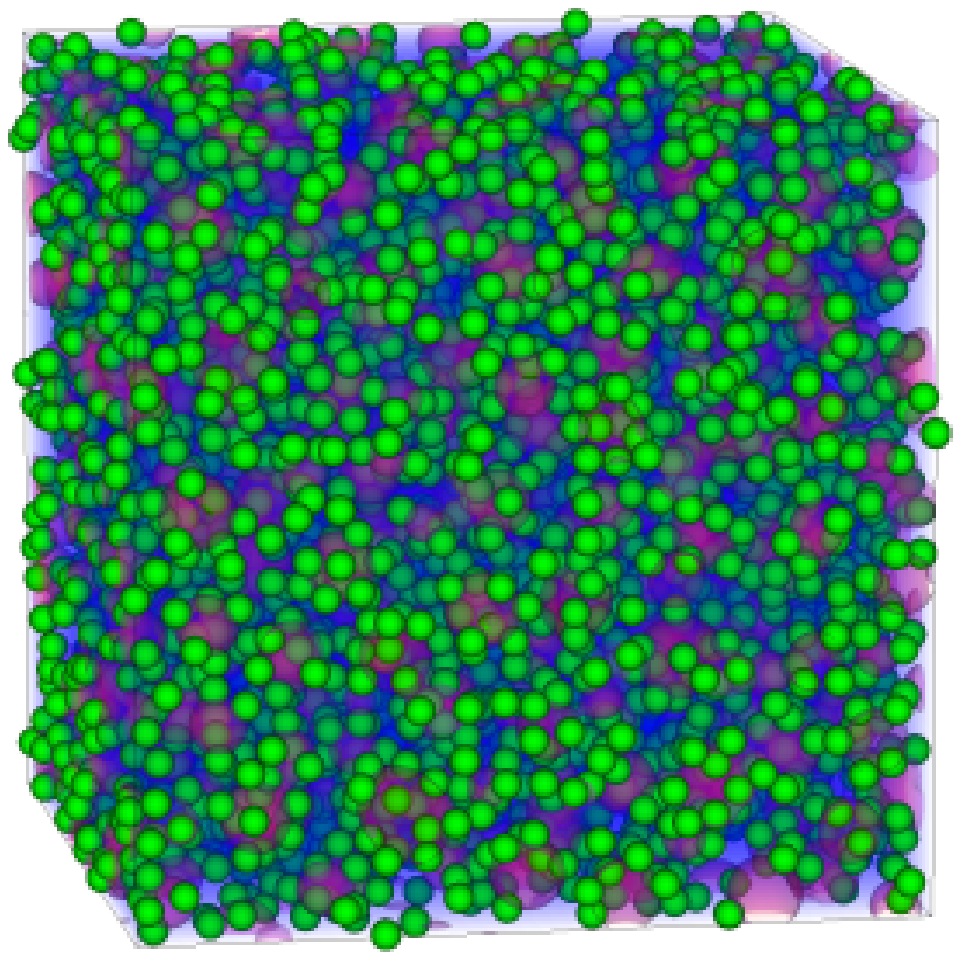} \\
$t=5000$ & $t=10000$ \\
\includegraphics[width=0.23\textwidth]{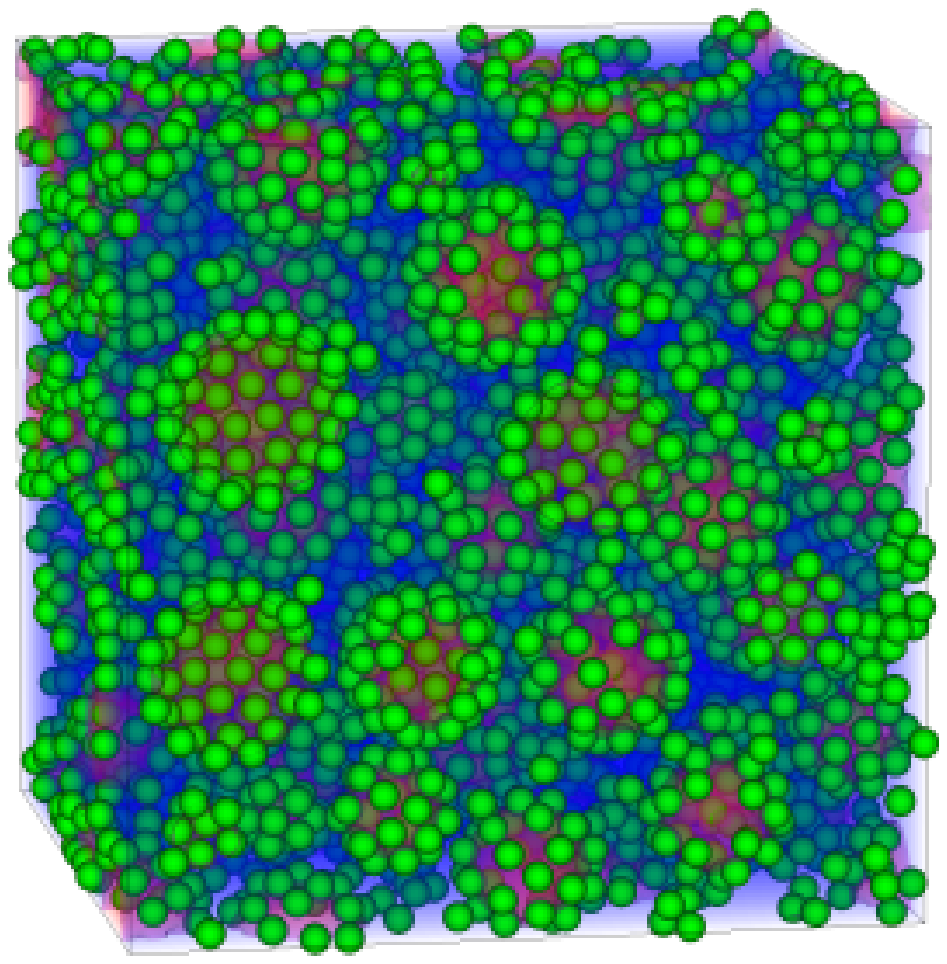} & 
\includegraphics[width=0.23\textwidth]{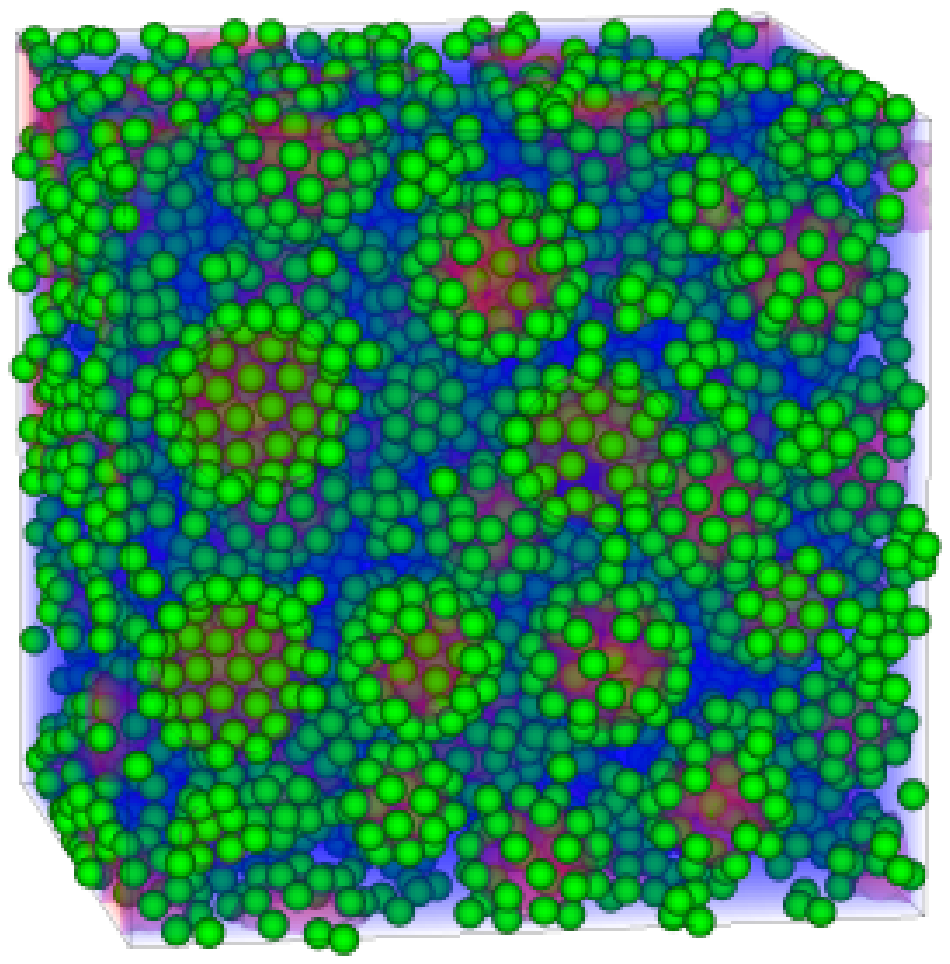}\\
$t=250000$ & $t=300000$ \\
\end{tabular}
\caption{(Color online) 3D visualisation of the system described in
Fig.~\ref{fig_pickering_c0.15-L}. The particles (green/light gray) and the fluids
(red/medium gray and blue/dark grey) are shown. The particles are attached to the interface
between the fluid components and the red fluid forms spherical droplets
inside the continuous blue fluid. While the change from timesteps 5000 to
10000 is significant the droplet growth is almost at rest between $t=2.5
\cdot 10^5$ and $t=3.0 \cdot 10^5$.}
\label{fig_pickering_images}
\end{figure}
Another well known phenomenon that can be observed in mixtures of
immiscible fluids and particles are Pickering
emulsions~\cite{bib:ramsden-1903,bib:pickering-1907}. Here, the particles
are not necessarily equally wettable by the fluids anymore. Also, the
ratio of the amount of fluid of different species deviates from 1. The
result is a system where one phase is continuous while the other forms
droplets which are stabilized by the
particles. The particles prevent the droplets from merging when they
collide and therefore stop the growth of the average droplet size. As
before the droplet size can be measured utilizing $L(t)$ as shown in
Fig.~\ref{fig_pickering_c0.15-L}. Here, the lattice is $256^3$, the
interaction constant 
$g_\textnormal{br}=0.09$, and the initial fluid density
$\rho_\textnormal{init}=0.66$. The fluid ratio is 1:3 and the particles
with a color of $\Delta \phi = -0.01$ have a volume concentration of 15\%.
As for the previously presented ``bijels'' a rapid growth of $L(t)$ from 5
to over 25 lattice units can be observed during the first 25000 timesteps
of the simulation. The domain growth slows down to a slight decelerating
growth afterwards. This agrees qualitatively with experimental results by
Arditty et al.~\cite{bib:arditty-whitby-binks-schmitt-leal-calderon}.

\begin{figure}[htb]
\includegraphics[width=1\linewidth]{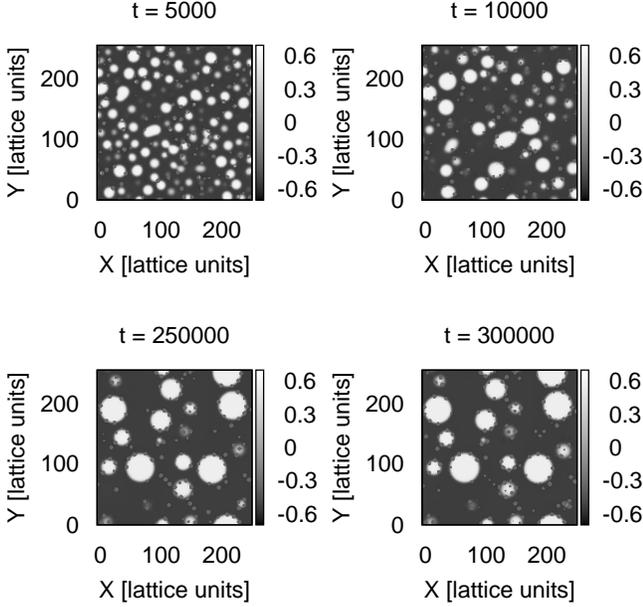}
\caption{2D cut at $z=0$ through the system described in figure
\ref{fig_pickering_c0.15-L} at different times. Shown is the order
parameter $\phi$.
}
\label{fig_pickering_colourfield}
\end{figure}
A three-dimensional visualisation of the order parameter $\phi$ and the
particles is shown in Fig.~\ref{fig_pickering_images} and accompanied by a two
dimensional cut of the system at $z=0$ in Fig.~\ref{fig_pickering_colourfield}.
The particle covered droplets as well as the slowing down of droplet growth can
be observed. While the system changes dramatically between timesteps 5000 and
10000 almost no difference can be observed when comparing step $2.5 \cdot 10^5$
to step $3.0 \cdot 10^5$.

\begin{figure}[htb]
\centerline{
\includegraphics[width=1\linewidth]{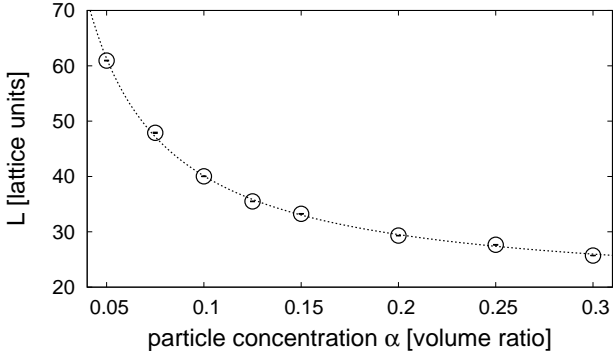}}
\caption{Average domain size after $1\cdot 10^6$ timesteps versus particle
concentration $\alpha$. System size $256^3$, $g_\textnormal{br}=0.08$,
$\rho_\textnormal{init}=0.7$, fluid ratio 1:3 and particle color $-0.01$.
Increasing $\alpha$ leads to a decreasing average domain
size. Also shown is a fit with equation
$\frac{2.12}{\alpha} + 18.91$. Error bars are given by the 
maximum deviation of $L_x$, $L_y$, $L_z$ from the mean.}
\label{fig_pickering_c}
\end{figure}
The influence of the interfacial area on the droplet size can be demonstrated
by modifying the particle concentration. The resulting average domain size for
different concentrations is shown in Fig.~\ref{fig_pickering_c}.  A higher
concentration leads to a larger stabilised interfacial area resulting in
smaller droplets: reducing the particle concentration from 0.15 to 0.05
corresponds to an increase of the average droplet size from 29 to 41 lattice
units. As the simulated system is finite, modifying the concentration of
particles does also change the volume of the two fluid components.  Therefore,
the inversely proportional relation between particle concentration and droplet
size as found by Arditty et
al.~\cite{bib:arditty-whitby-binks-schmitt-leal-calderon} does not apply here,
but has to be shifted by a constant offset.  $L=\frac{2.12}{\alpha}+ 18.91$ is
found to be a good fit of the data presented in Fig.~\ref{fig_pickering_c}.

\begin{figure}[htb]
\centerline{
\includegraphics[width=1\linewidth]{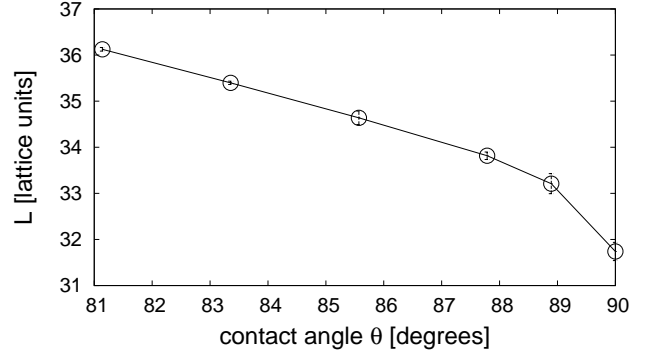}}
\caption{Average domain size after $1.5 \cdot 10^5$ timesteps versus contact
angle $\theta$. System size $256^3$, $g_\textnormal{br}=0.08$,
$\rho_\textnormal{init}=0.7$, fluid ratio 5:9 and particle concentration
$\alpha=0.15$. Neutrally wetting particles lead to a small $L$ with a large
deviation between the spatial directions while strongly colored particles lead
to a larger average domain size with almost no deviations. Error bars are given
by the maximum deviation of $L_x$, $L_y$, $L_z$ from the mean.}
\label{fig_pickering_colour}
\end{figure}
As expected the colour and thus the contact angle of the particle has a drastic
influence on the formed structure. While strongly colored particles with
contact angles different from 90 degrees lead to spherical droplets, neutrally
wetting particles result in droplets that are not as spherical anymore. We
observe structures that are similar to the ones found by Kim et
al.~\cite{bib:kim-stratford-adhikari-cates} for their simulation of neutrally
wetting particles. These structures are extended in one of the directions. This
results in a reduction of the measured average domain size $L$, while the
difference between the directions increases. This difference is expressed
through the error bar in Fig.~\ref{fig_pickering_colour}. The values of the
contact angle shown in the figure are obtained from the mapping presented in
Fig.~\ref{fig_kw_f}b.

\section{Transition from Bijel to Pickering emulsion}
\begin{figure}[htb]
\centerline{
\includegraphics[width=1\linewidth]{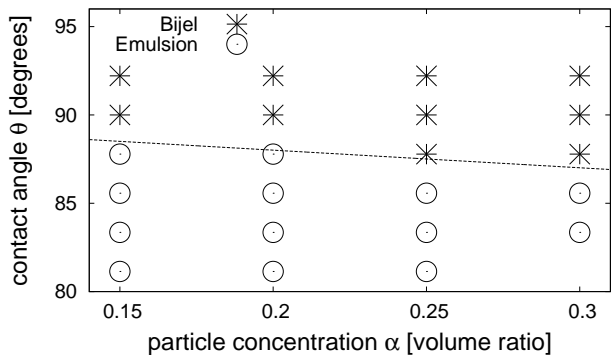}}
\caption{System state in dependence on the contact angle and concentration
after $3\cdot 10^4$ timesteps. The system size is $128^3$,
$\rho_\textnormal{init}=0.7$, and the ratio of the two fluid species is kept
fixed at 5:9. For contact angles larger than 90 degrees the system always
relaxes towards a bijel, while for strongly negative coloring a Pickering
emulsion is obtained. The line is a guide to the eye.\label{fig_pickering_c-c}
}
\end{figure}

\begin{figure}[htb]
\centerline{
\includegraphics[width=1\linewidth]{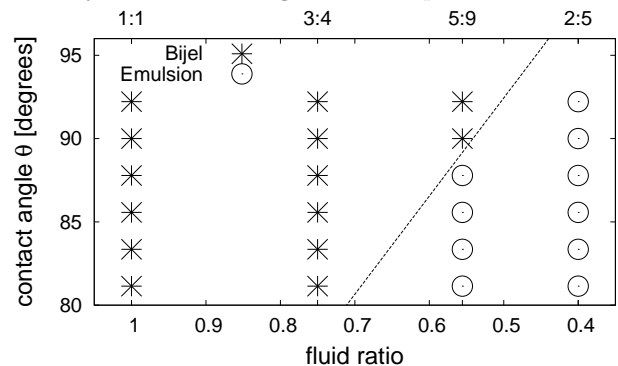}}
\caption{System state in dependence on the contact angle and the fluid ratio
after $3\cdot 10^4$ timesteps. The system size is $128^3$,
$\rho_\textnormal{init}=0.7$, and the particle concentration is kept fixed at
0.2. For fluid ratios between 1:1 and 3:4 also for contact angles larger than
90 degrees a bijel is obtained. For larger concentration ratios it depends on
the particle color or contact angle if a bijel or a Pickering emulsion is
produced. The line is a guide to the eye.
\label{fig_pickering_c-ratio}}
\end{figure}

In the current section it is demonstrated how the contact angle, the
particle volume concentration and the ratio of the two fluid species
determine the final state of the system to be a bijel or a Pickering
emulsion. Phase diagrams depending on the various simulation parameters
are presented in Fig.~\ref{fig_pickering_c-c} and
\ref{fig_pickering_c-ratio}.
In order to reduce the computational cost, the size of the lattice has
been reduced to 128$^3$ in this section. However, by performing a small
number of 512$^3$ sized simulations it has been confirmed that finite size
effects are still below an acceptable limit and do not influence the final
physical state of the system. If the system categorizes as bijel or
Pickering emulsion is determined visually after $3\cdot 10^4$ timesteps.
$\rho_\textnormal{init}$ is kept fixed at 0.7 and $g_{br}$ is set to 0.08
in all simulations.

Figure~\ref{fig_pickering_c-c} shows a phase diagram in dependence on the
particle concentration and the contact angle (see Fig.~\ref{fig_kw_f}b for the
mapping between the particle color and the contact angle). The ratio of the two
fluid species is kept fixed at 5:9. For contact angles larger than 90 degrees
the system always relaxes towards a bijel, while for strongly negative coloring
and thus smaller contact angles a Pickering emulsion is obtained. The particle
concentration, however, only has a minor influence on the final state. It can
only be noticed that for small concentrations the formation of a Pickering
emulsion is more favored for smaller contact angles. The line is only a guide
to the eye since the exact position of the transition from bijel to Pickering
emulsion would require substantially more data points. 

In Fig.~\ref{fig_pickering_c-ratio} the final system state is depicted in
dependence on the contact angle and the fluid ratio. A fluid ratio of at
least 3:4 results also for contact angles larger than 90 degrees in a bijel, while
for a fluid ratio of 2:5 even neutrally wetting particles are able to
stabilize a Pickering emulsion.
As already shown in Fig.~\ref{fig_pickering_c-c} for a fluid ratio of 5:9
it depends on the contact angle if the system relaxes towards a bijel or
a Pickering emulsion. This behavior can be explained by the interplay
between interface curvature and interface size: it is only energetically
beneficial if the work required to maintain a curved interface
is not larger than the cost due to the increased size of the interface,
where the latter can be overcome by a higher concentration of particles at
the interface.

\section{Conclusion} 
\label{sec_conclusion}
In this paper we proposed a new method allowing the simulation of particles
with variable contact angle in multicomponent fluid flows.  We have studied the
influence of the model parameters on the resulting fluid-particle interactions
and shown that our approach is able to simulate the formation of ``bijels'' and
Pickering emulsions. By computing phase diagrams we have demonstrated how the
transition from bijel to Pickering emulsion is determined by the contact angle
between particle and fluids, the particle concentration, and the ratio of the
two fluid species: while the wettability of the particles and the fluid
ratio strongly influence the transition from a bijel to a Pickering state,
the particle volume concentration only has a minor impact.
 
\begin{acknowledgments}
We like to thank the DFG for funding within SFB 716. Further funding is
acknowledged from FOM (IPP IPoGII) and NWO/STW (VIDI grant of J.~Harting). We
thank F. Janoschek and S. Schmieschek for fruitful discussions. In particular
the contribution of F. Janoschek to the development of the simulation code is
highly acknowledged. Simulations have been performed at the Scientific
Supercomputing Centre Karlsruhe and the J\"ulich Supercomputing Center. 
\end{acknowledgments}


\end{document}